\RequirePackage{lineno}
\documentclass[floatfix,twocolumn,preprintnumbers,amsmath,amssymb,prl,superscriptaddress]{revtex4} 
\usepackage{url}
\usepackage{graphicx}
\usepackage{dcolumn}
\usepackage{xspace}
\usepackage{bm}
\usepackage{soul}
\usepackage{ulem}
\usepackage{xcolor}
\usepackage{pdfpages}

\begin{document}

\title{Probing (Hyper)Nuclei Wave Functions and Production Mechanisms in $\sqrt{s_{\rm{NN}}}=200$ GeV Isobar Collisions at RHIC}
\author{The STAR Collaboration}

\begin{abstract}

The study of nuclei and hypernuclei production is a powerful tool to investigate the formation mechanism of loosely bound states in high-energy heavy-ion collisions. A key prediction from coalescence models is a strong suppression of the hypertriton (${}^{3}_{\Lambda}\rm{H}$) compared to ${}^{3}\rm{He}$ production in small collision systems due to the larger radius of ${}^{3}_{\Lambda}\rm{H}$. In this letter, the STAR collaboration reports measurements on (hyper)nuclei ($^{3}_{\Lambda}\rm{H}, {}^{3}_{\bar{\Lambda}}\rm{\bar{H}}, {}^{3}\rm{He}, {}^{3}\rm{\overline{He}}, t$) production at mid-rapidity in Ru+Ru and Zr+Zr collisions at $\sqrt{s_{\rm{NN}}}=200$ GeV as a function of collision centrality. We find that the ratios $N_{t}N_{p}/N_{d}^{2}$ and $S_{3} =(N_{{}^{3}_{\Lambda}\rm{H}}/N_{{}^{3}_{}\rm{He}})/(N_{\Lambda}/N_{p})$ deviate significantly from thermal model expectations. Coalescence calculations incorporating realistic non-Gaussian wave functions for the $d$ and ${}^{3}_{\Lambda}\rm{H}$ provide an improved description of the data, while Gaussian descriptions of the ${}^{3}_{\Lambda}\rm{H}$ wave function fail to simultaneously reproduce the measured $S_{3}$ and the binding energy of ${}^{3}_{\Lambda}\rm{H}$. These results suggest that the ${}^{3}_{\Lambda}\rm{H}$ wave function contains larger short-distance $d$--$\Lambda$ probability than implied by a Gaussian ansatz, demonstrating the potential of heavy-ion production measurements as a probe of hypernuclear wave functions and the underlying hyperon--nucleon interaction.

\end{abstract}


\maketitle
Production of light nuclei and their antiparticles in relativistic heavy-ion collisions has been studied since the early 1970s~\cite{Nagamiya:1981sd, Gutbrod:1976zzr, Gosset:1976cy,E814:1994kon, E864:2000auv, E886:1994ioj}. In the early 2000s, such studies were expanded to include (anti)hypernuclei, bound states of hyperons and nucleons~\cite{STAR:2021orx, STAR:2023fbc, STAR:2010gyg, ALICE:2015oer, E864:2002xhb,Bellini:2020cbj}. However, despite considerable experimental~\cite{STAR:2022hbp,ALICE:2022veq,E864:2002xhb,STAR:2010gyg,ALargeIonColliderExperiment:2021puh, STAR:2022fnj} and theoretical~\cite{Andronic:2010qu,Reichert:2022mek,Scheibl:1998tk,Steinheimer:2012tb, Aichelin:2019tnk,Zhou:2025zgn,Wang:2023gta,ALICE:2025byl,HADES:2025qum} efforts, our understanding of the production mechanisms of light nuclei and hypernuclei remains incomplete. 

Two classes of production models are commonly used. In thermal models, all particles, including (hyper)nuclei, are assumed to be in equilibrium at chemical freeze-out, while the mechanism by which loosely bound states are formed and survive in the hadronic medium remains under active discussion~\cite{Braaten:2024cke}. Within this framework, their yields are determined by the freeze-out temperature, baryochemical potential, and volume, and are largely insensitive to nuclear structure~\cite{Andronic:2010qu}. In coalescence models, nuclei are formed when the spatial coordinates and momenta of the constituent nucleons are close to each other~\cite{Reichert:2022mek}. In more sophisticated coalescence 
approaches based on the Wigner-function formalism~\cite{Scheibl:1998tk}, the formation probability is determined by the overlap between the phase-space distribution of the nucleon emitting source and the nuclear wave function. Recent state-of-the-art studies have shown that deuteron production can be sensitive to the choice of the nuclear wave function~\cite{Mahlein:2023fmx}, indicating that production yields may probe the internal structure of loosely bound nuclei.

This sensitivity is expected to become particularly important in small collision systems, such as peripheral heavy-ion, proton–nucleus, or proton–proton collisions, where the nucleon emitting source is significantly smaller than in central heavy-ion collisions. When the source radius becomes comparable to or smaller than the nuclear radius, the overlap between the source and the nuclear wave function becomes strongly suppressed and increasingly sensitive to the spatial structure of the nucleus, effectively allowing the wave function to be “resolved” through yield measurements~\cite{Mahlein:2023fmx,Bellini:2020cbj, Mahlein:2025bla, Leung:2025jwe}.

The hypertriton ($^{3}_{\Lambda}\rm{H}$), a bound state of $p$, $n$, and $\Lambda$, provides an especially sensitive test case because of its small $\Lambda$ binding energy. The measured world-average value, $B_{\Lambda}=0.163\pm0.036~\mathrm{MeV}$~\cite{Eckert:2022dyz, Kasagi:2025mvh, ALICE:2022sco, STAR:2019wjm, Chaudhari1968, Juric:1973zq, Mayeur1966ADO, ammar1962, Prakash1961OnTB}, corresponds to a root-mean-square $d$--$\Lambda$ separation of $\langle r_{d\Lambda}\rangle=9.8^{+1.1}_{-0.7}~\mathrm{fm}$~\cite{Leung:2025jwe,Liu:2024ygk}, larger than the typical nucleon source size in heavy-ion collisions. As a result, coalescence models predict a strong suppression of $^{3}_{\Lambda}\rm{H}$ production in small systems compared to thermal models~\cite{Zhang:2009ba, Sun:2018mqq}. Measurements across different system sizes can therefore provide stringent tests of the production mechanism and potentially constrain the wave function of the $^{3}_{\Lambda}\rm{H}$. Such constraints are important for understanding hyperon–nucleon interactions, which play a key role in determining the equation of state of neutron stars~\cite{Gerstung:2020ktv}. Detailed measurements of the $^{3}_{\Lambda}\rm{H}$ yield in large systems are available from RHIC~\cite{STAR:2023fbc, STAR:2026khp} and LHC~\cite{ALICE:2024koa, ALICE:2015oer}. Yet, measurements in small systems remain scarce.

In this letter, we report measurements of (anti)hypernuclei $^{3}_{\Lambda}\rm{H}( {}^{3}_{\bar{\Lambda}}\rm{\bar{H}})$, (anti)nuclei ${}^{3}\rm{He}({}^{3}\rm{\overline{He}})$, and $t$ yields at mid-rapidity in Ru+Ru and Zr+Zr (hereafter abbreviated as isobar) collisions at $\sqrt{s_{\rm{NN}}}=200$ GeV. Different yield ratios will be explored as a function of multiplicity to investigate the system size dependence of hypernuclei and nuclei production. The analyses are based on minimum-bias (MB) data collected with the STAR detector in 2018. The MB data were triggered by the coincidence of both vertex position detectors (VPDs)~\cite{Llope:2014nva} located at forward and backward pseudorapidity. The time projection chamber (TPC)~\cite{Anderson:2003ur} is used to reconstruct charged-particle tracks within the pseudorapidity range $|\eta|<1$. The vertex position along the beam axis ($V_{z,\rm{TPC}}$) and its radial distance from the beam axis ($V_{r}$) of each event are reconstructed with the TPC tracks, and are required to be $-35<V_{z,\rm{TPC}}<25$ cm and $V_{r}<2$ cm, to reject contamination from interactions with the beam pipe~\cite{STAR:2021mii}. The VPDs also provide vertex position information along the beam direction ($V_{z,\rm{VPD}}$). The difference between $V_{z,\rm{TPC}}$ and $V_{z,\rm{VPD}}$ is required to be less than 5 cm to suppress pile-up events~\cite{STAR:2021mii}. The collision centrality is determined by comparing the measured charged-particle multiplicity within $|\eta|<0.5$ to a Glauber model simulation~\cite{STAR:2021mii,Miller:2007ri}. For our analyses, we select $0–80\%$ centrality MB events, resulting in a total of 3.6 billion events that satisfy our selection criteria. 

\begin{figure}[ht]
	\centering	
	\includegraphics[width=0.8\linewidth]{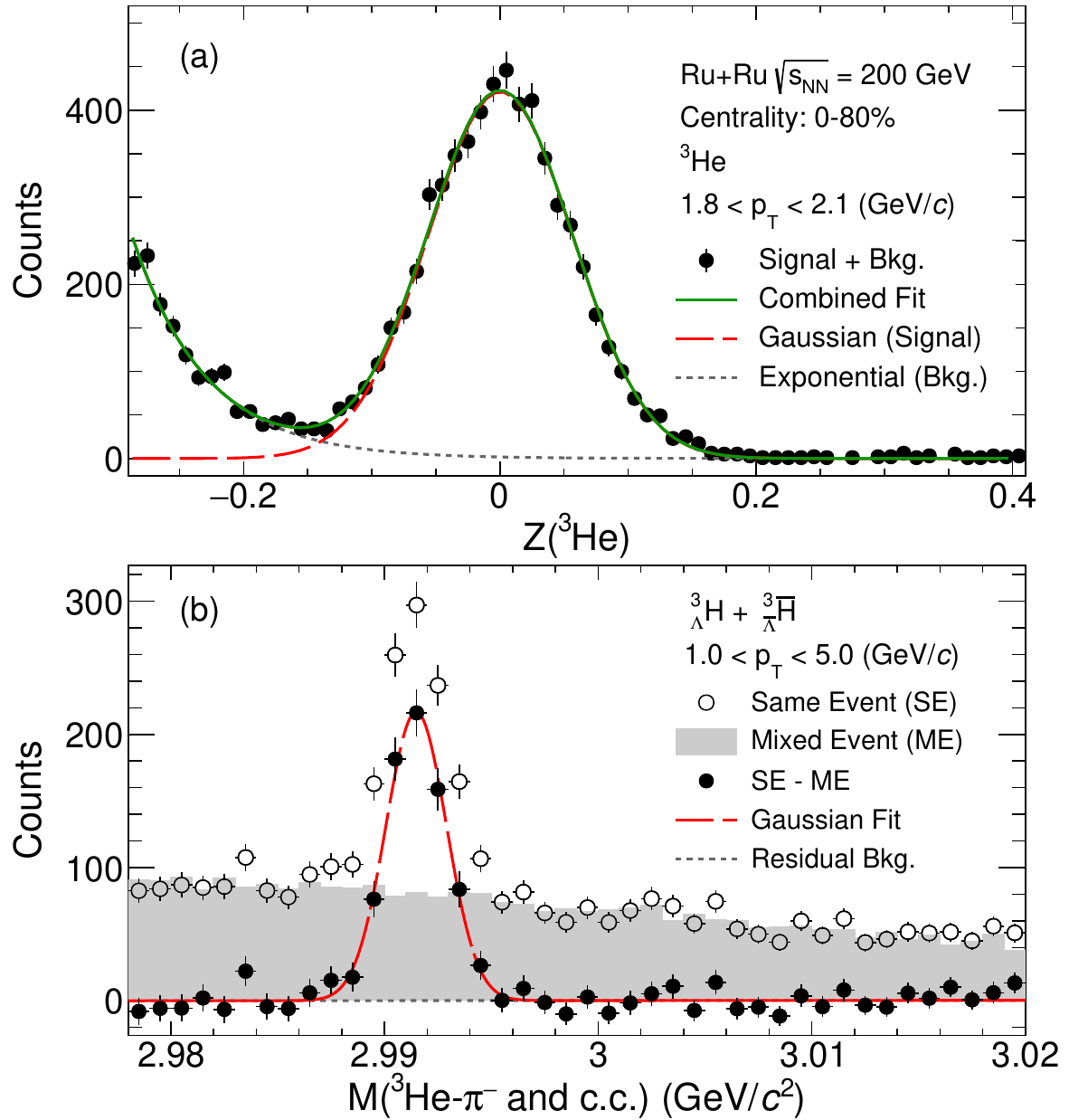}
    \caption{(a) $Z$ distribution of ${}^{3}\rm{He}$ candidates with $p_{T} \in [1.8,2.1]$ GeV$/c$. The black circles represent the data, the red and gray lines represent Gaussian and exponential functions modeling the signal and background, respectively. The green line is the sum of the two functions. (b) The invariant mass distribution of ${}^{3}\rm{He}$-$\pi^{-}$ and ${}^{3}\rm{\overline{He}}$-$\pi^{+}$ pairs. The black open circles represent the data, gray shaded region represents the background. The black solid circles represent the background-subtracted data. The gray dashed line represents a linear function describing the residual background, while the red line represents the sum of the Gaussian function describing the signal and the residual background. }
    \label{fig:pid}
\end{figure}

For the analysis of nuclei, reconstructed tracks are required to have a distance of closest approach to the primary collision vertex of less than 1 cm
and have at least 20 hit points measured in the TPC to ensure track quality. To avoid double-counting, each track is required to have more than $52\%$ of the maximum possible hit points. The identification of $\mathrm{^{3}He}(\mathrm{^{3}\overline{He}})$ and $t$ is performed using
information from the TPC and Time-of-Flight (TOF) detectors~\cite{Llope:2003ti}. For $\mathrm{^{3}He}(\mathrm{^{3}\overline{He}})$, particle identification is performed by comparing the measured ionization energy loss ($dE/dx$) in the TPC with the theoretical expectation ($dE/dx_{\text{Bichsel}}$)~\cite{Bichsel:2006cs}, using the variable $Z = \ln \left( \langle dE/dx\rangle \,/\, \langle dE/dx\rangle_{\text{Bichsel}} \right)
$. Particle candidates are selected within a $\pm3\,\sigma$ window around the peak of the $Z$ distribution. Figure~\ref{fig:pid}(a) shows the $Z$ distribution of ${}^{3}\rm{He}$ candidates within $p_{T}\in[1.8,2.1]$ GeV$/c$. The background distributions are modeled using exponential functions, and raw yields are obtained by integrating the background-subtracted distributions. For $t$, additional TOF information is used due to the lower signal-to-background ratio. The particle mass squared is calculated via ${(m/q)}^{2}={(p/q)}^{2}\left(1 / {\beta}^{2}-1\right)$, where $q$ is the charge of the track, $p/q$ is the rigidity measured by the TPC, and $\beta$ is the track velocity measured by the TOF. Similarly, the raw yields are extracted by integrating the background-subtracted ${(m/q)}^{2}$ distribution within a $\pm3\,\sigma$ window around the signal peak.  

Hypernuclei $^{3}_{\Lambda}{\rm{H}}(^{3}_{\bar{\Lambda}}\rm{\bar{H}})$ are reconstructed via weak decay channels, i.e.
$^{3}_{\Lambda}\rm{H}(^{3}_{\bar{\Lambda}}\rm{\bar{H}}) \rightarrow \mathrm{^{3}He}(\mathrm{^{3}\overline{He}}) + \pi^{\pm}$, using the KFParticle package~\cite{Ju:2023xvg} which utilizes a Kalman Filter algorithm to optimize the estimation of decay topology and kinematic variables. The daughter particle tracks ($\mathrm{^{3}He}(\mathrm{^{3}\overline{He}})$ and $\pi^{\pm}$) are identified using the $dE/dx$ measured by the TPC. The combinatorial backgrounds are constructed by the mixed-event method~\cite{Kornakov:2018uzv}. For the selection of hypernuclei candidates, the topological variables and their correlations are studied using a machine‑learning method: an extreme‑gradient‑boosting decision tree (XGBDT)~\cite{Chen:2016btl}. The XGBDT is trained with the signal samples from a Monte-Carlo (MC) simulation of the STAR detector (detailed in next paragraph) and the background samples from the mixed events. The XGBDT training procedure was validated using independent training and testing samples, and no evidence of overtraining was found. The hypernuclei candidates are selected based on the XGBDT output value, which characterizes the probability for a candidate to be a real signal, to optimize the statistical significance. Figure~\ref{fig:pid}(b) shows the invariant mass distributions of ${}^{3}\rm{He}$-$\pi^{-}$ and ${}^{3}\rm{\overline{He}}$-$\pi^{+}$ pairs with $1<p_{T}<5$ GeV$/c$ . In addition to subtracting the combinatorial background, a linear fit using the side-band region is performed to remove any residual background. Raw yields are extracted with a counting method after subtracting the backgrounds~\cite{STAR:2026khp}. 

After the raw yields are obtained for each particle in $p_{T}$ and centrality bins ($0-10\%$, $10-20\%$, $20-40\%$, $40-80\%$) within the rapidity range $|y|<$ 0.8 for $^{3}_{\Lambda}\rm{H}(^{3}_{\bar{\Lambda}}\rm{\bar{H}})$ and $|y|<$ 0.5 for nuclei, the invariant yields are calculated using the equation:
\begin{equation}
    \frac{d^{2}N}{2\pi p_{\rm{T}^{}} dp_{\rm{T}}dy} = \left(\frac{1}{\rm{B.R.}}\right)\frac{1}{2\pi p^{\rm{}}_{\rm{T}} \Delta p_{\rm{T}}\Delta y}\frac{\Delta N^{\rm{raw}}}{N^{\rm{event}}A\epsilon}, 
\end{equation}
where \rm{B.R.} is the branching ratio of the analyzed decay channel (applicable only for the hypernuclei analysis) and is estimated to be $23\pm3\%$~\cite{STAR:2026khp}, $\Delta p_{\rm{T}}$ and $\Delta y$ are the widths of the transverse momentum bin and rapidity interval, $N^{\rm{event}}$ is the number of analyzed events, $\Delta N^{\rm{raw}}$ is the raw yield, and $A\epsilon$ is the acceptance and reconstruction efficiency which includes corrections for the TPC acceptance, track reconstruction, and particle identification. Acceptance and tracking efficiencies are calculated using an embedding technique in which the TPC response to MC particles (and decay daughters) is simulated in the STAR detector described in \textsc{Geant3}~\cite{Brun:1119728}. Since \textsc{Geant3} does not properly account for (anti)nucleus absorption in detector materials, we evaluate the absorption of $^{3}\rm{He}$($\mathrm{^{3}\overline{He}}$) and $t$ using \textsc{Geant4}. The loss of (anti)nuclei due to interactions
with the detector material within \textsc{Geant3} was scaled to
match the values from \textsc{Geant4}~\cite{STAR:2001pbk}. The magnitude of this correction is less than $3\%$($6\%$) for nuclei(antinuclei). The particle identification efficiency corrections, which account for the efficiency of the TOF detector in identifying tracks measured by the TPC, are estimated using data-driven methods~\cite{STAR:2017sal}. Finally, we also subtract the weak decay contributions from the nuclei yields in order to determine their primordial yields. For $^{3}\rm{He}$($\mathrm{^{3}\overline{He}}$) and $t$, decay contributions from $^{3}_{\Lambda}\rm{H}(^{3}_{\bar{\Lambda}}\rm{\bar{H}}) \rightarrow \pi^{\mp}+{ }^3 \mathrm{He}(\mathrm{^{3}\overline{He}})$ and ${ }_{\Lambda}^3 \mathrm{H} \rightarrow \pi^{0}+t$ are considered. Using the aforementioned embedding technique~\cite{STAR:2022hbp,STAR:2019bjj}, these contributions are estimated to be $0$--$4\%$, depending on $p_{T}$. The corrected $p_{T}$ spectra can be found in Ref.~\cite{SupplementalMaterial}. $p_{T}$-integrated yields are calculated by extrapolating to the unmeasured $p_{T}$ regions with individual Blast-Wave~\cite{Schnedermann:1993ws} fits to the $p_{T}$ spectra. 


The four major sources of systematic uncertainties on the $p_{T}$-integrated yields are the TPC tracking efficiency, extrapolation to the unmeasured $p_{T}$ regions, mismatch of topological variables between data and MC for $^{3}_{\Lambda}\rm{H}$ reconstruction, and the $^{3}_{\Lambda}\rm{H}$ branching ratio. For the TPC tracking efficiency, the track quality selection cuts are varied in addition to an assignment of $5\%$ uncertainty per track~\cite{STAR:2024lvy}, which amounts to $\sim$$8\%$($13\%$) for nuclei(hypernuclei) analysis. The systematic uncertainty on extrapolation is estimated using different functions~\cite{SupplementalMaterial}, and is $4$--$24\%$($9$--$22\%$) for nuclei(hypernuclei) analysis. For the hypernuclei reconstruction efficiency, the selection cut on XGBDT output value is varied. In addition, the lifetime assumption of the ${}^{3}_{\Lambda}\rm{H}$ in the simulations is varied within a $\pm1\,\sigma$ window of the average experimental lifetime~\cite{Eckert:2022dyz}. These variations contribute an uncertainty of $\sim$$6\%$. Finally, a $13\%$ uncertainty is assigned to the ${}^{3}_{\Lambda}\rm{H}$ branching ratio~\cite{STAR:2026khp}. Systematic uncertainties are discussed further in Ref.~\cite{SupplementalMaterial}. Different sources of systematic uncertainties are assumed to be uncorrelated and added in quadrature.

Nuclei and hypernuclei yields are heavily influenced by the nucleon and hyperon yields. In order to compare their production from different collision systems, we focus on two special ratios, $N_{t}N_{p}/N_{d}^{2}$ and $S_{3}$, defined as
\[
S_{3} =
\frac{(N_{{}^{3}_{\Lambda}\mathrm{H}}+N_{{}^{3}_{\bar{\Lambda}}\bar{\mathrm{H}}})/(N_{{}^{3}\mathrm{He}}+N_{{}^{3}\overline{\mathrm{He}}})}
{(N_{\Lambda}+N_{\bar{\Lambda}})/(N_{p}+N_{\bar{p}})}.
\]

\noindent These two ratios share a common feature: the denominator and numerator share the same baryon and isospin number, thus, these two ratios are, to first-order, insensitive to baryon density and isospin effects, and the associated theoretical uncertainties largely cancel~\cite{Zhang:2009ba, Sun:2017xrx}. The nuclei and hypernuclei yields in this study are combined with the $p$, $d$, and $\Lambda$ yields from Refs.~\cite{placeholder,STAR:2024lvy} to construct these ratios.

\begin{figure}[ht!]                     
    \centering           \includegraphics[width=0.99\linewidth]{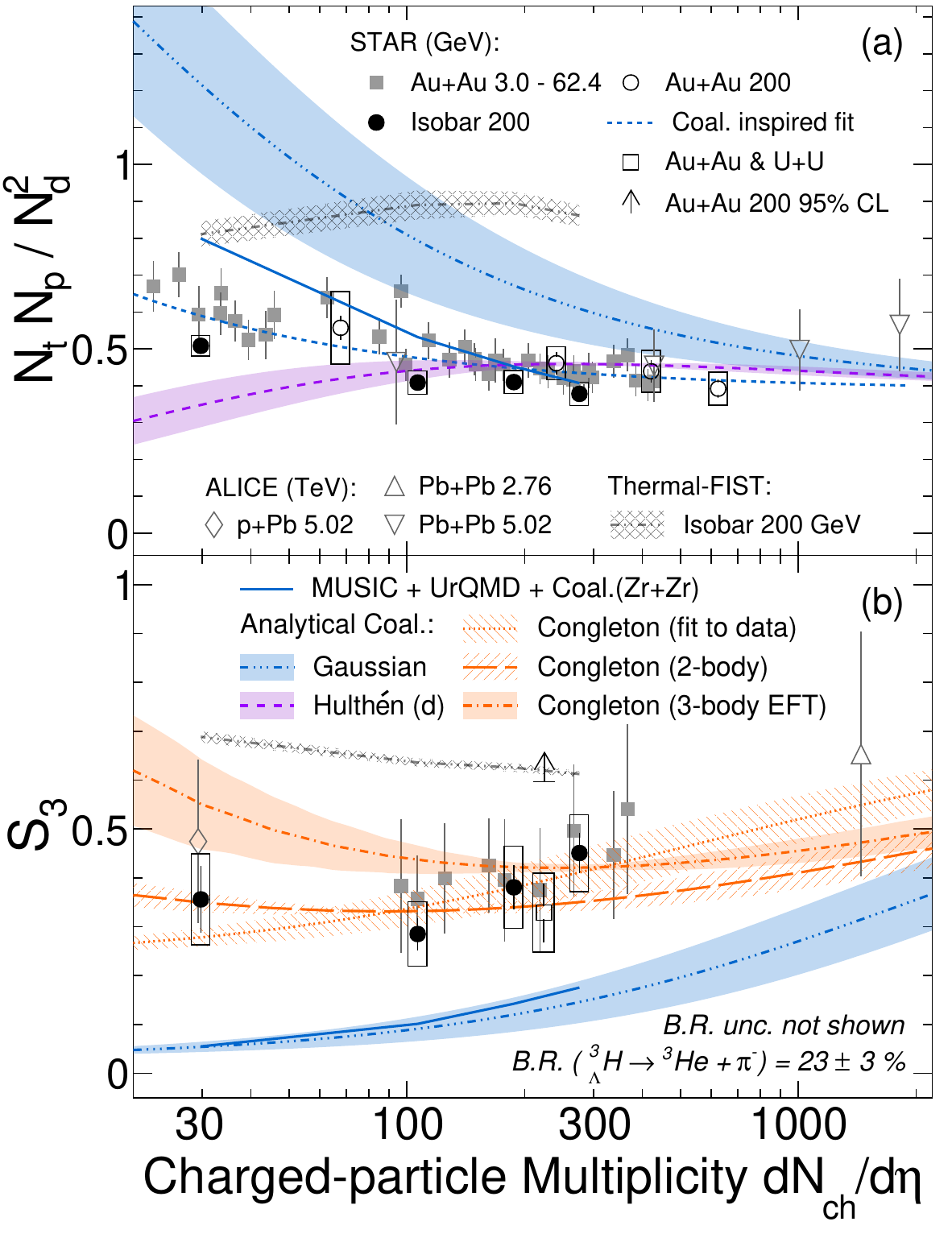}
    \caption{The multiplicity dependence of (a) $N_{t}N_{p}/N_{d}^{2}$ and (b) $\mathrm{S}_{3}$. Results from isobar collisions are indicated using black circles, and are compared to results from other collision systems~\cite{STAR:2022hbp,ALICE:2013mez,ALICE:2015wav,ALICE:2022veq, STAR:2010gyg,ALargeIonColliderExperiment:2021puh, ALICE:2015oer, STAR:2023fbc, ALICE:2019fee, ALICE:2019bnp, STAR:2026khp}. For $\sqrt{s_{\rm{NN}}}=200$ GeV data, the vertical lines and boxes are the statistical and systematic
    uncertainties, respectively, while the vertical lines indicate the sum of statistical and systematic
    uncertainties for other data. The uncertainty on the ${}^{3}_{\Lambda}\rm{H}$ B.R., fully correlated among all data points, is not shown. The blue dashed line in the upper panel shows the coalescence-inspired fit to the STAR data~\cite{STAR:2022hbp}. The blue dot-dashed line, purple dashed line, and orange lines correspond to analytical coalescence calculations employing different wave functions~\cite{Bellini:2020cbj,Leung:2025jwe}, while the blue solid line represents the hybrid coalescence calculation~\cite{Zhao:2021dka}. The gray lines show thermal-model calculations~\cite{Vovchenko:2019pjl}. The bands indicate model uncertainties.}
    
\label{fig:H3L_ratio}    
\end{figure}

Figure~\ref{fig:H3L_ratio}(a) presents the ratio $N_{t}N_{p} / N_{d}^{2}$ as a function of charged particle multiplicity ($dN_{\rm{ch}}/d\eta$), which has been shown to be a good proxy for the size of the nucleon emitting source~\cite{ALICE:2025wuy,Lisa:2005dd}. The $p_{T}$ spectra of $\pi^{\pm}$, $K^{\pm}$, and $p(\bar{p})$~\cite{STAR:2024lvy} are used to estimate $dN_{\rm{ch}}/d\eta$ for each centrality bin. Results from isobar collisions at $\sqrt{s_{\mathrm{NN}}}$ = 200 GeV are compared with those from other collision energies and systems~\cite{STAR:2022hbp,ALICE:2013mez,ALICE:2015wav,ALICE:2022veq}. All data points, independent of collision energy and system, approximately follow a common trend that decreases mildly with multiplicity.
Our results are compared with predictions from a thermal model. Thermal-FIST~\cite{Vovchenko:2019pjl} calculations, constrained by hadron yield measurements in isobar collisions~\cite{STAR:2024lvy}, are shown as a gray dash-dotted line, with the band indicating the model uncertainty (see Ref.~\cite{SupplementalMaterial} for details). The calculations significantly overestimate the isobar data ($\chi^{2}/\mathrm{NDF}=374.3/4$).

We also compare our results with three coalescence approaches: the MUSIC+UrQMD hybrid model~\cite{Zhao:2021dka}, an analytical model~\cite{Bellini:2020cbj}, and a coalescence-inspired fit to the STAR data~\cite{STAR:2022hbp}. In all frameworks, the nucleus formation probability is determined by the overlap of the phase-space distributions of the nucleon source with the nuclear wave function. The primary difference between these approaches lies in their treatment of the phase-space distributions of the nucleon source. In the hybrid model, the distributions are obtained from dynamical simulations combining the hydrodynamic model MUSIC~\cite{Denicol:2018wdp, Shen:2017bsr}, which describes the evolution of the quark–gluon plasma, and the transport model UrQMD~\cite{Bleicher:1999xi}, which models the subsequent hadronic transport stage. In contrast, the analytical model and the coalescence fit approximate the nucleon source with a Gaussian distribution characterized by a radius parameter $R_{\rm inv}$. In the analytical model, $R_{\rm inv}$ is taken from a data-driven parameterization extracted from femtoscopy measurements~\cite{ALICE:2025wuy}, while for the coalescence fit, $R_{\rm inv}$ is modelled as $R_{\rm inv}=p_{1}\times (dN_{\rm{ch}}/d\eta)^{1/3}$, where $p_{1}$ is determined via a fit to the $N_{t}N_{p}/N_{d}^{2}$ data~\cite{SupplementalMaterial}.

For computational convenience, the nuclear wave functions are commonly approximated by Gaussian forms. Using this approximation for $d$ and $t$, all three coalescence approaches predict a decreasing trend with multiplicity. While the same hybrid approach describes the Au$+$Au data over $\sqrt{s_{\rm NN}}=7.7$–$200$~GeV well~\cite{Zhao:2021dka}, the corresponding calculation for Zr$+$Zr collisions at $\sqrt{s_{\rm NN}}=200$~GeV overestimates the isobar data at low multiplicity. The analytical approach shows a similar overprediction, with $\chi^{2}$/NDF$=25.1/4$. Meanwhile, the coalescence fit 
can reproduce the data by allowing the source size parameter to vary freely. However, this additional 
flexibility may distort the physical source size~\cite{SupplementalMaterial}, and the agreement with the data does not necessarily imply a more realistic description of the underlying production process.

While a Gaussian wave function is a reasonable approximation for ${}^{3}\mathrm{He}$ and  $t$~\cite{Bellini:2020cbj}, the $d$ has a comparatively broader wave function, making $d$ production more susceptible to the details of the wave function~\cite{Mahlein:2023fmx}. We therefore also employ the Hulthén wave function~\cite{Hulthen1957}, using the same parameterization as Ref.~\cite{Bellini:2020cbj}, within the analytical coalescence framework. The Hulthén wave function provides an improved description ($\chi^{2}$/NDF=14.4/4) relative to the Gaussian, but underestimates the yields at $dN_{\rm{ch}}/d\eta<80$. This deviation may reflect limitations of the analytical model, particularly the assumption of a Gaussian nucleon source that neglects nucleon momentum correlations~\cite{Pratt:1997pw}, together with uncertainties associated with the choice of Hulthén-wave-function parameters~\cite{Lamia:2012zz}, both of which may become increasingly important in small systems.

 Figure~\ref{fig:H3L_ratio}(b) shows the ratio $S_3$ as a function of $dN_{\rm{ch}}/d\eta$. The results are compared with measurements from other collision systems and energies~\footnote{For the $\sqrt{s_{NN}}=7.7$--$27$ GeV measurements, we show the ratio $(^{3}_{\Lambda}\rm{H}/t)/(\Lambda/p)$ instead, due to the lack of ${}^{3}\rm{He}$ data.}~\cite{STAR:2010gyg,ALargeIonColliderExperiment:2021puh, ALICE:2015oer, STAR:2023fbc} and predictions from thermal and coalescence models. The $S_{3}$ values from all systems and energies are consistent with a common trend and show no significant multiplicity dependence. Thermal model calculations tend to overpredict the measured $S_{3}$ values across the centrality intervals, corresponding to $\chi^{2}/\mathrm{NDF}=22.9/4$.

Using a Gaussian wave function for ${}^{3}_{\Lambda}\mathrm{H}$ constrained to reproduce the world-average binding energy, corresponding to $\langle r_{d\Lambda}\rangle=9.8^{+1.1}_{-0.7}~\mathrm{fm}$, both the hybrid and analytical coalescence models underestimate the isobar data across all multiplicity intervals, yielding $\chi^{2}/\mathrm{NDF}=13.9/4$ and $14.9/4$, respectively. The absence of the strong suppression expected for such a large-radius Gaussian wave function indicates that this ansatz does not provide an adequate description of the ${}^{3}_{\Lambda}\mathrm{H}$ wave function. To investigate this further, we employ the Congleton wave function within the analytical coalescence framework. The Congleton wave function is derived by treating the ${}^{3}_{\Lambda}\rm{H}$ as a two-body system governed by a $d$--$\Lambda$ interaction, and takes the form
$\phi(q) \propto \exp(-q^{2}/Q_{\Lambda}^{2})/(q^{2}+\alpha_{\Lambda}^{2})$~\cite{Congleton:1992kk}, where $q$ is the relative $d$--$\Lambda$ momentum. 
We consider two parameter sets for $(Q_{\Lambda},\alpha_{\Lambda})$: one obtained from the original two-body calculation,
$(1.17~\mathrm{fm}^{-1},\,0.068~\mathrm{fm}^{-1})$~\cite{Congleton:1992kk},
and another motivated by an effective-field-theory (EFT) treatment that includes three-body forces,
$(2.5~\mathrm{fm}^{-1},\,0.068~\mathrm{fm}^{-1})$~\cite{Hildenbrand:2019sgp}. Both parameter sets yield $\langle r_{d\Lambda}\rangle\approx10~\mathrm{fm}$, similar to the Gaussian case. In contrast to the Gaussian wave function, the Congleton parameterization describes the data well for both the two-body and three-body  parameter sets, yielding $\chi^{2}/\mathrm{NDF}=1.6/4$ and $4.9/4$, respectively.

\begin{figure}[ht!]         
    \centering    
    \includegraphics[width=0.47\linewidth]{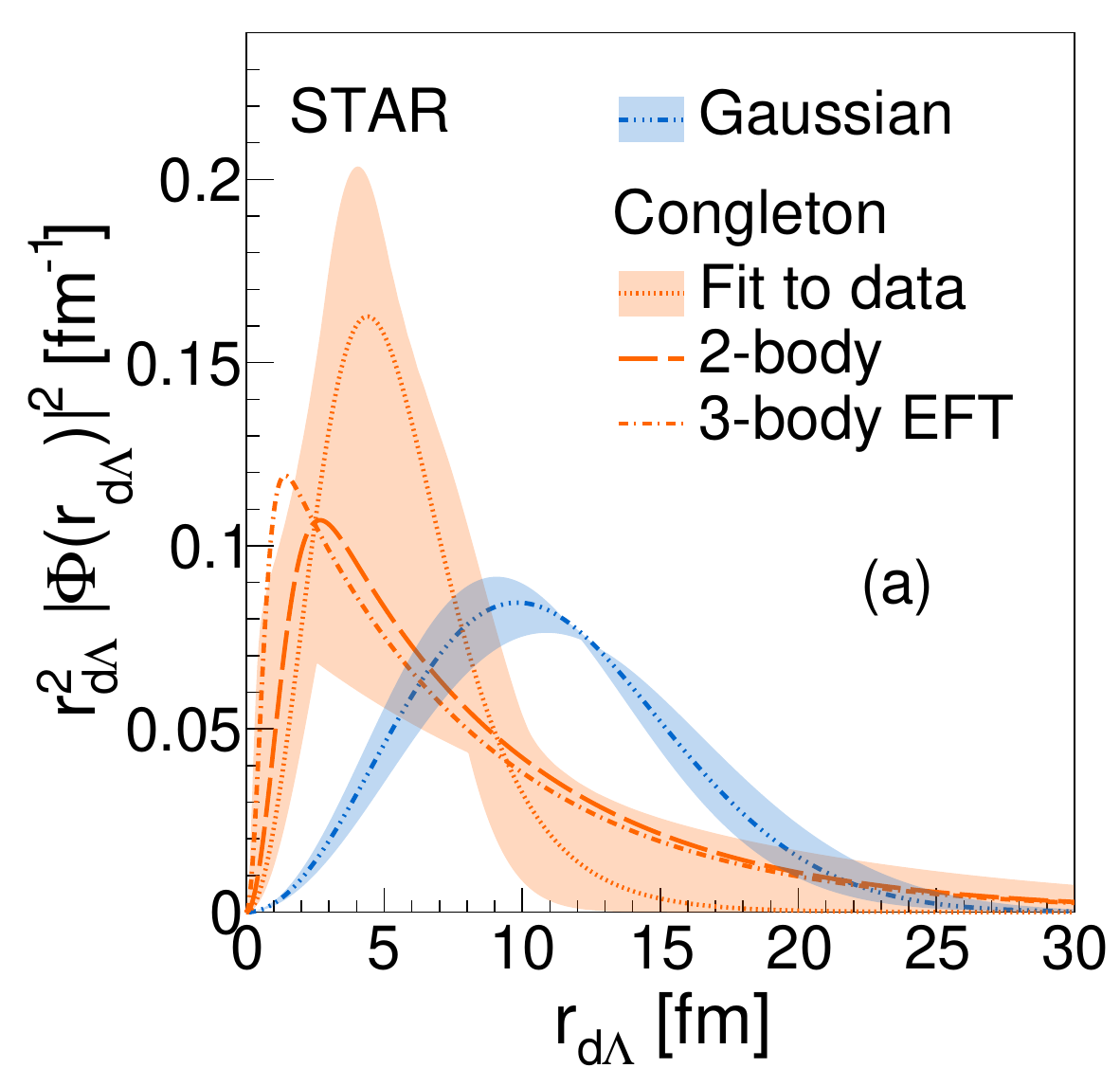}
    \includegraphics[width=0.50\linewidth]{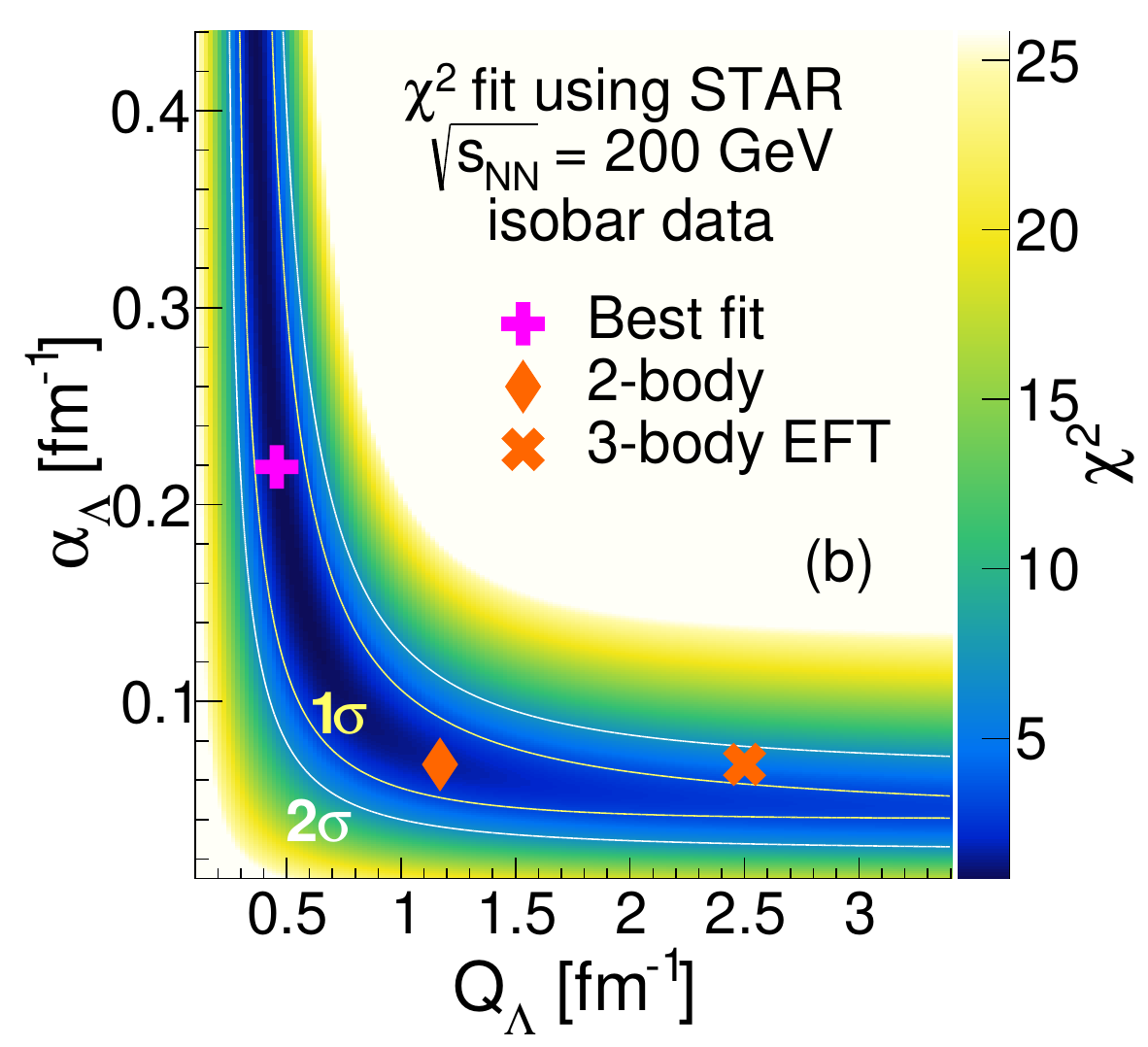}
    \caption{(a) Radial probability distributions $r_{d\Lambda}^{2}|\Phi(r_{d\Lambda})|^{2}$ obtained from the fit, compared with the corresponding two-body Congleton, three-body EFT Congleton, and Gaussian wave functions. The orange and blue bands denote the $1\sigma$ confidence interval from the fit and the uncertainty associated with the binding-energy constraint for the Gaussian wave function respectively. (b) $\chi^{2}$ distribution in the $(Q_{\Lambda},\alpha_{\Lambda})$ parameter space obtained from the $\chi^{2}$ fit to the isobar data using the Congleton wave function. The yellow and white contours represent the $1\sigma$ and  $2\sigma$ confidence regions respectively. The markers indicate the parameters corresponding to the best fit (yellow cross), two-body (blue diamond) and three-body EFT (blue diagonal cross) parameters.}
    \label{fig:unconstrained_chi2_scan}    
\end{figure}

Figure~\ref{fig:unconstrained_chi2_scan}(a) shows the radial probability distributions $r_{d\Lambda}^{2}|\Phi(r_{d\Lambda})|^{2}$ for the Gaussian wave function and the Congleton wave function using the two-body and three-body EFT parameter sets, where $\Phi(r_{d\Lambda})$ denotes the ${}^{3}_{\Lambda}\rm{H}$ wave function. Although all three wave functions correspond to $\langle r_{d\Lambda}\rangle\approx10~\mathrm{fm}$, the two Congleton wave functions exhibit substantially larger probability at small $r_{d\Lambda}$. Since coalescence is sensitive to the overlap between the wave function and the nucleon source, whose characteristic size in isobar collisions is $1$--$3~\mathrm{fm}$~\cite{SupplementalMaterial}, the enhanced short-distance probability leads to a larger predicted $S_{3}$ and a significantly improved description of the data. These results indicate that the isobar $S_{3}$ measurements are primarily sensitive to the short-distance structure of the ${}^{3}_{\Lambda}\rm{H}$ wave function rather than its overall size.

To further study the constraints imposed by the data on the ${}^{3}_{\Lambda}\mathrm{H}$ wave function within the analytical coalescence framework, we perform a $\chi^{2}$ fit to the isobar data using the Congleton parameterization with $Q_{\Lambda}$ and $\alpha_{\Lambda}$ treated as free parameters. The best-fit values, $(Q_{\Lambda},\alpha_{\Lambda})=(0.46~\mathrm{fm}^{-1},0.22~\mathrm{fm}^{-1})$, also provide an excellent description of the measurements as shown in Fig~\ref{fig:H3L_ratio}(b). Figure~\ref{fig:unconstrained_chi2_scan}(b) shows the resulting $\chi^{2}$ distribution. The $1\,\sigma$ confidence regions form an L-shaped structure, with one branch extending toward small $\alpha_{\Lambda}$ at larger $Q_{\Lambda}$ and another toward larger $\alpha_{\Lambda}$ at small $Q_{\Lambda}$. Within this region, the extracted $\langle r_{d\Lambda}\rangle$ spans $6.3^{+11.1}_{-1.3}~\mathrm{fm}$, corresponding to a binding energy $B_{\Lambda}=0.52^{+0.53}_{-0.48}~\mathrm{MeV}$. This range is consistent with the world-average value, although the uncertainty is large because the two-parameter Congleton form allows a broad family of wave-function shapes. Nevertheless, the corresponding radial probability distributions, shown by the orange band in Fig.~\ref{fig:unconstrained_chi2_scan}(a), consistently exhibit enhanced probability at small $r_{d\Lambda}$ relative to the Gaussian ansatz. This indicates that the ${}^{3}_{\Lambda}\mathrm{H}$ wave function contains a larger short-distance $d$--$\Lambda$ probability than implied by a Gaussian ansatz.

Finally, we perform a fit to the isobar data using a Gaussian wave function with the width treated as a free parameter. The fit provides a good description of the $S_{3}$ data, with $\chi^{2}_{\mathrm{min}}=0.9$, and yields $\langle r_{d\Lambda}\rangle =4.69^{+0.56}_{-0.44}~\mathrm{fm}$, corresponding to $B_{\Lambda}=1.28^{+0.51}_{-0.38}~\mathrm{MeV}$. This result is incompatible with the world-average value at the $4.2\,\sigma$ level~\cite{SupplementalMaterial}, indicating that although a Gaussian ansatz can reproduce the isobar data, it does so only by requiring an unrealistically compact ${}^{3}_{\Lambda}\mathrm{H}$ wave function.

In summary, we report measurements of ${}^{3}_{\Lambda}\mathrm{H}$(${}^{3}_{\bar{\Lambda}}\bar{\mathrm{H}}$), ${}^{3}\mathrm{He}$(${}^{3}\overline{\mathrm{He}}$), and $t$ production in $\sqrt{s_{\mathrm{NN}}}=200$ GeV isobar collisions. The measured ratios $N_{t}N_{p}/N_{d}^{2}$ and $S_{3}$ deviate significantly from thermal-model calculations, while coalescence calculations incorporating realistic nuclear wave functions provide an improved description of the data. In particular, Gaussian descriptions of the ${}^{3}_{\Lambda}\mathrm{H}$ wave function fail to simultaneously reproduce the measured $S_{3}$ ratio and the binding energy of ${}^{3}_{\Lambda}\mathrm{H}$. In contrast, the Congleton wave function containing  enhanced short-distance $d$--$\Lambda$ probability, successfully describes both constraints. Because the spatial structure of ${}^{3}_{\Lambda}\mathrm{H}$ is determined by the underlying hyperon–nucleon interaction, these results demonstrate that hypernuclei production yields in heavy-ion collisions can serve as a sensitive probe of hyperon–nucleon interactions relevant for hypernuclear structure and neutron-star matter.

We thank the RHIC Operations Group and SDCC at BNL, the NERSC Center at LBNL, and the Open Science Grid consortium for providing resources and support.  This work was supported in part by the Office of Nuclear Physics within the U.S. DOE Office of Science, the U.S. National Science Foundation, National Natural Science Foundation of China, Chinese Academy of Science, the Ministry of Science and Technology of China and the Chinese Ministry of Education, NSTC Taipei, the National Research Foundation of Korea, Czech Science Foundation and Ministry of Education, Youth and Sports of the Czech Republic, Hungarian National Research, Development and Innovation Office, New National Excellency Programme of the Hungarian Ministry of Human Capacities, Department of Atomic Energy and Department of Science and Technology of the Government of India, the National Science Centre and WUT ID-UB of Poland, German Bundesministerium f\"ur Bildung, Wissenschaft, Forschung and Technologie (BMBF), Helmholtz Association, Ministry of Education, Culture, Sports, Science, and Technology (MEXT), Japan Society for the Promotion of Science (JSPS), and Agencia Nacional de Investigacion y Desarrollo de Chile (ANID), Chile.

\bibliography{references} 


\clearpage
\onecolumngrid  


\section*{Supplementary material}

# Supplementary Material: Probing (Hyper)Nuclei Wave Functions and Production Mechanisms in $\sqrt{s_{\rm NN}} = 200$ GeV Isobar Collisions at RHIC

The STAR Collaboration

(Dated: July 30, 2026)

## (HYPER)NUCLEI $p_{\rm T}$ DISTRIBUTIONS

Figure 1 shows the $p_{\rm T}$ spectra for $t$, $^3$He, $^3\overline{\rm He}$ in the rapidity range $|y| < 0.5$, and $({}^{3}_{\Lambda}{\rm H}+{}^{3}_{\bar\Lambda}\bar{\rm H})\,/\,2$ in the rapidity range $|y| < 0.8$ in 0-10%, 10-20%, 20-40%, and 40-80% Ru+Ru and Zr+Zr collisions at $\sqrt{s_{\rm NN}} = 200$ GeV. The spectra are fitted with Blast-Wave functions:

$$\frac{d^2N}{dp_{\rm T}dy} \propto p_{\rm T} \int_0^R r dr {\rm m_T} I_0 \left(\frac{p_{\rm T}{\rm sinh}\rho}{T_{\rm kin}}\right) K_1 \left(\frac{{\rm m_T}{\rm cosh}\rho}{T_{\rm kin}}\right), \quad (1)$$

where $I_0$ and $K_1$ are modified Bessel functions, $\rho = \tanh^{-1}\beta$, $\beta = \beta_{\rm s}(r/R)^n$, $\beta_{\rm s}$ denotes the surface velocity, $r/R$ is the relative radial position in the thermal source, $n$ denotes the flow velocity profile exponent (fixed to unity for this study) and $T_{\rm kin}$ denotes the kinetic freeze-out temperature. The fitted Blast-Wave function is used to extrapolate to the unmeasured $p_{\rm T}$ region and obtain the $p_{\rm T}$-integrated yields $dN/dy$. For the nuclei $dN/dy$, the extrapolated contribution from the unmeasured $p_{\rm T}$ region is 43–73%, depending on centrality, while for the combined ${}^{3}_{\Lambda}$H and ${}^{3}_{\bar\Lambda}\bar{\rm H}$ $dN/dy$, the contribution is 30–70%.

Due to limited statistics, the ${}^{3}_{\Lambda}{\rm H}({}^{3}_{\bar\Lambda}\bar{\rm H})$ analysis is performed using the combined Ru+Ru and Zr+Zr data set. The $t$ and $^3$He ($^3\overline{\rm He}$) yields are combined from the two systems with equal weights, assuming fully correlated systematic uncertainties. The $p$ and $\Lambda$ yields from Ref. [1] are treated in the same manner.

## SYSTEMATIC UNCERTAINTIES

For the $p_T$ spectra, we consider six sources of systematic uncertainties. The TPC tracking efficiency and signal extraction procedure affect both the nuclei and hypernuclei analyses. For the hypernuclei analysis, additional uncertainties arise from the mismatch of topological variables between data and MC (including the XGBDT output constructed from topological variables and the assumed ${}^{3}_{\Lambda}$H lifetime in the MC) and the ${}^{3}_{\Lambda}$H branching ratio (B.R.). For the nuclei analysis, uncertainties from feed-down corrections are also considered. For the $dN/dy$, extrapolation to the unmeasured $p_T$ regions is also considered. In the following, we discuss the systematic uncertainties of $dN/dy$.

The uncertainties on the TPC tracking efficiency, the TPC track quality selection cuts are varied in addition to an assignment of 5% base uncertainty per track [2], which amounts to ∼8%(13%) for nuclei(hypernuclei) analysis. The uncertainty on signal extraction for the nuclei analysis is estimated by comparing the bin-counting and template-fit methods, resulting in a systematic uncertainty of < 1%. For the hypernuclei analysis, the signal-extraction uncertainty is estimated by varying the bin-counting window and by propagating the covariance matrix of the residual-background fit to the estimated background contribution in the signal region, resulting in a total uncertainty of 1–4%. For the hypernuclei reconstruction efficiency, the selection cut on XGBDT output value is varied. In addition, the lifetime assumption of the ${}^{3}_{\Lambda}$H in the simulations is varied within a $\pm 1\,\sigma$ window of the average experimental lifetime. $\tau({}^{3}_{\Lambda}{\rm H}) = 228 \pm 12$ ps is used for this analysis [3]. These variations contribute an uncertainty of 3–5%. Finally, the uncertainty on the feed-down correction is evaluated by varying the ${}^{3}_{\Lambda}$H spectra within their uncertainties and propagating the effect to the feed-down contribution, resulting in an uncertainty of less than 1%.

Different sources of systematic uncertainties are assumed to be uncorrelated and added in quadrature. The extrapolation uncertainties are estimated by alternate choices of the fitting functions, including:

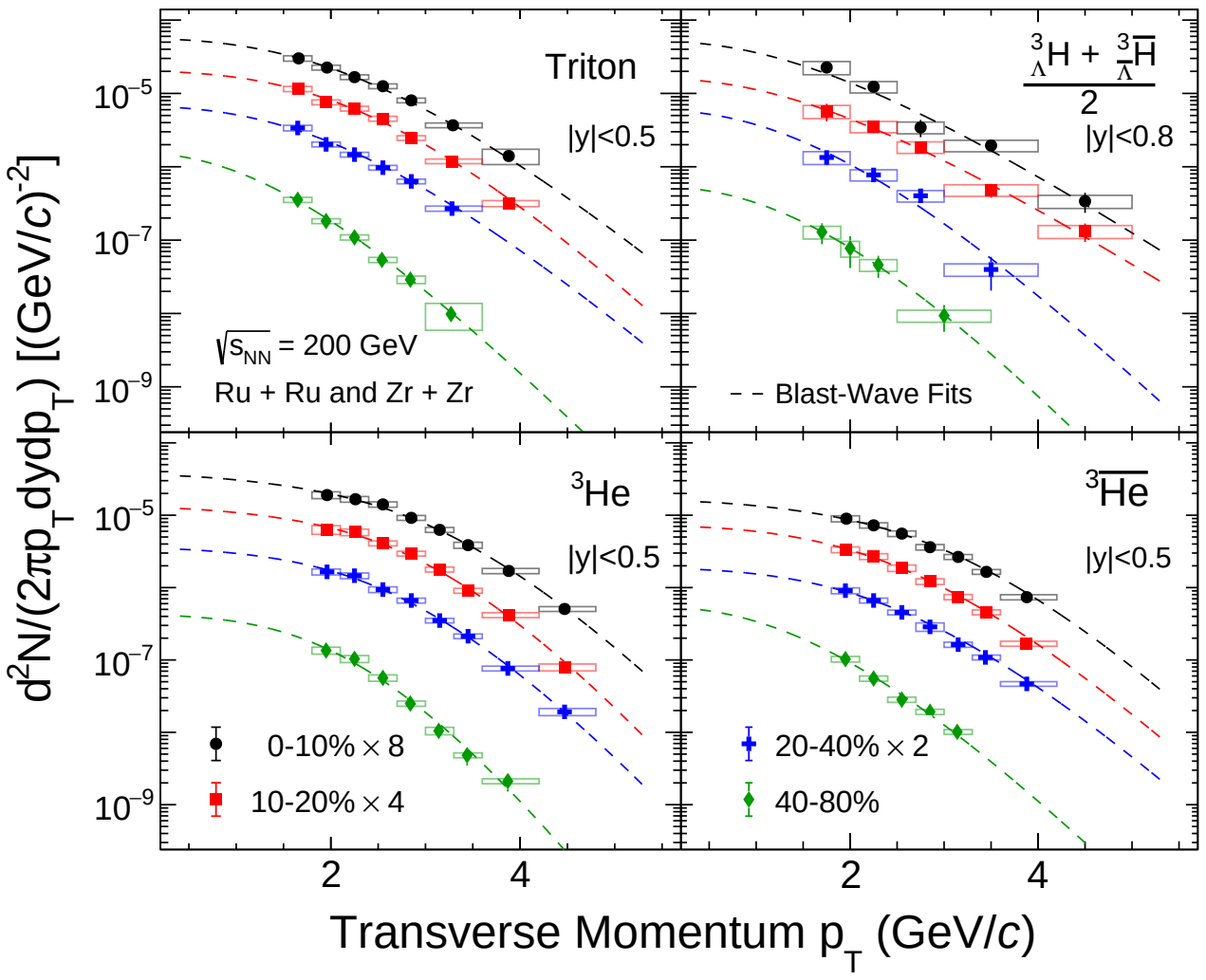


FIG. 1. Transverse momentum ($p_T$) spectra of $t$, $^3$He, $^3\overline{\rm He}$, and $({}^{3}_{\Lambda}{\rm H} + {}^{3}_{\bar\Lambda}\bar{\rm H})\,/\,2$ in different centrality bins in isobar collisions at $\sqrt{s_{\rm NN}} = 200$ GeV. The spectra are scaled by different factors for clarity of presentation. Systematic uncertainties are represented by boxes. The dotted lines are fits using the Blast-Wave model.

- Boltzmann: $\frac{d^2N}{dp_T dy} \propto p_T m_T e^{-m_T/T}$,
- $m_T$-exponential: $\frac{d^2N}{dp_T dy} \propto p_T e^{-m_T/T}$,
- $p_T$-guassian : $\frac{d^2N}{dp_T dy} \propto p_T e^{-(p_T/T)^2}$, and
- Lévy: $\frac{d^2N}{dp_T dy} \propto p_T \frac{(n-1)(n-2)}{nT(nT+m(n-2))} \left(1+\frac{m_T}{nT}\right)^{-n}$,

where $m_T = \sqrt{m^2+p_T^2}$, $m$ denotes the particle mass, $T$ and $n$ are free parameters constrained by fitting to the data. The Lévy function is not applied to the $^3_\Lambda$H spectra due to its poor fitting performance. Table I summarizes the systematic uncertainties of $dN/dy$ for the different particles.

TABLE I. The relative systematic uncertainties of dN/dy

| | $t$ | $^3$He | $^3\overline{\text{He}}$ | $^3_\Lambda\text{H} + ^3_{\bar\Lambda}\bar{\text{H}}$ |
|---|---|---|---|---|
| Signal extraction | $< 1\%$ | $< 1\%$ | $< 1\%$ | $1-4\%$ |
| Tracking efficiency | 5–10% | 6–10% | 6–8% | 10–15% |
| Topological cuts | – | – | – | 2–4% |
| $^3_\Lambda$H lifetime | – | – | – | 3–5% |
| Feed-down | $< 1\%$ | $< 1\%$ | $< 1\%$ | – |
| Extrapolation | 4–24% | 13–23% | 5–18% | 9–22% |
| $^3_\Lambda$H B.R. | – | – | – | 13% |
| Total | 6–25% | 17–29% | 8–24% | 20–24% |

When propagating the systematic uncertainties from the particle $dN/dy$ measurements to the $N_tN_p/N_d^2$ ratio, the 5% base tracking-efficiency uncertainty per track and the uncertainty associated with the extrapolation to the unmeasured kinematic region are treated as correlated, while the remaining sources of uncertainty are propagated assuming no correlations. The comparison to theoretical model calculations is quantified using a $\chi^2$ test of the form $(\text{data}-\text{theory})^2/(\sigma^2_{\text{data}}+\sigma^2_{\text{theory}})$. Since the $^3_\Lambda$H B.R. uncertainty is fully correlated among different data points, it is incorporated using a covariance-matrix approach.

## THERMAL MODEL CALCULATIONS

The thermal model calculations shown in the main manuscript are obtained using the Thermal-FIST package (v1.5) [4]. The ideal hadron resonance gas model is applied using the Boltzmann approximation, assuming the system reaches full chemical equilibrium with the electric ($\gamma_q$) and strangeness ($\gamma_s$) suppression factors fixed to 1. The canonical ensemble is applied, requiring the net conserved charges to be conserved across the correlation volume $V_c$ in each microscopic state. The net strangeness number ($\langle S\rangle$) is fixed to zero, while $\langle B\rangle$ is configured according to the Ref. [2]. $\langle Q\rangle$ is estimated as $Z/A * \langle B\rangle$, where $Z/A$ is the ratio of atomic and mass number of $^{96}_{44}$Ru. The volume $V_c$ is fixed to $V_c = \frac{dV}{dy}$. The only free parameters are the chemical freeze-out temperature $T_{\text{ch}}$ and the volume $V$. By fitting the primordial yields of light hadrons including $\pi^\pm, K^\pm, p(\bar{p}), \Lambda(\bar{\Lambda})$ and $\Xi^\pm$ [1, 2], we obtain the thermodynamic parameters and predict the (hyper)nuclei yields. To assess the uncertainties, The following variations are considered: $\langle Q\rangle = Z/A*\langle B\rangle$ or 0, $V_c = \frac{dV}{dy}$ or $3\frac{dV}{dy}$ and $T_{\text{ch}}$ by its uncertainty ($T_{\text{ch}}$ or $T_{\text{ch}} \pm \sigma$). The uncertainties of our predictions are estimated by adding the model differences between the default and the varied configurations in quadrature. Tab. II lists the Thermal-FIST calculations on the yield ratios $N_tN_p/N_d^2$ and $S_3$ in different centrality classes of the $\sqrt{s_{\text{NN}}} = 200$ GeV isobar collisions.

TABLE II. Thermal-FIST calculations on the yield ratios $N_tN_p/N_d^2$ and $S_3$ in different centrality classes of $\sqrt{s_{\text{NN}}} = 200$ GeV isobar collisions.

| | 0-10% | 10-20% | 20-40% | 40-80% |
|---|---|---|---|---|
| $\frac{N_tN_p}{N_d^2}$ | $0.862 \pm 0.043$ | $0.895 \pm 0.047$ | $0.890 \pm 0.062$ | $0.811 \pm 0.183$ |
| $S_3$ | $0.612 \pm 0.009$ | $0.626 \pm 0.015$ | $0.636 \pm 0.012$ | $0.689 \pm 0.049$ |

## COALESCENCE INSPIRED FIT

Following Refs. [5, 6], a thermally equilibrated static spherical Gaussian nucleon source is assumed, with the source radius parameterized as $R_{\text{inv}} = p_1\,(dN_{\text{ch}}/d\eta)^{1/3}$. Within this framework, the ratio $N_tN_p/N_d^2$ can be expressed as

$$\frac{N_tN_p}{N_d^2} = p_0\left(\frac{R^2_{\text{inv}}+\frac{2}{3}r_d^2}{R^2_{\text{inv}}+\frac{1}{2}r_t^2}\right)^3, \quad (2)$$

where $r_d = 1.96$ fm and $r_t = 1.59$ fm are the root-mean-square radii of deuteron and triton, respectively. A simultaneous fit to all STAR data gives $p_0 = 0.38 \pm 0.007$ and $p_1 = 0.81 \pm 0.043$, and the resulting $R_{\text{inv}}$ vs. $dN_{\text{ch}}/d\eta$ relationship is shown in Fig. 2.

## ANALYTICAL COALESCENCE MODEL CALCULATIONS

For the analytical coalescence model calculations presented in the main manuscript, we adopt the framework established in Ref. [7], and our specific implementation of this framework is further discussed in Ref. [8]. The key equations relating ($B_2$, $B_{3(^3\text{He})}$, $B_{3\Lambda}$) with the source size $R_{\text{inv}}$ in Ref. [7] are used. Since our analysis employs proton yields corrected for weak-decay feed-down, the chaoticity parameters for $t$ and $^3$He are fixed to unity. In contrast, the $\Lambda$ yields are not corrected for feed-down from the decay $\Sigma^0 \to \Lambda + \gamma$. Thermal model calculations estimate this feed-down contribution to be about 27%.

Owing to the long lifetime of the $\Sigma^0$, we adopt a chaoticity parameter of $1-0.27=0.73$ for the ${}^3_\Lambda$H calculations, as discussed in Ref. [8].

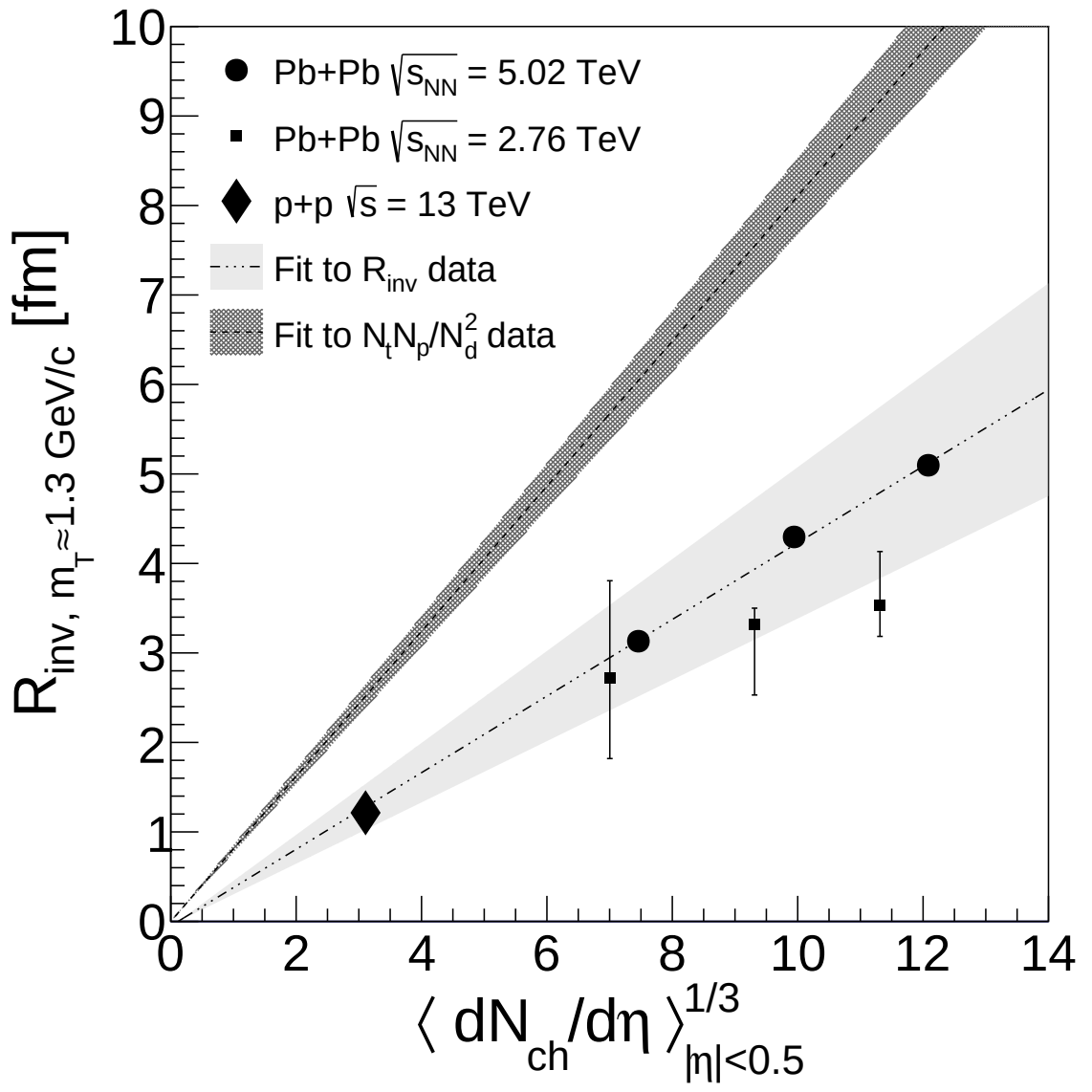


FIG. 2. The femtoscopic source size of protons at $m_T \approx 1.3$ GeV/$c$ from $p+p$ [9] and Pb+Pb [10, 11] collisions at LHC. The red band shows the parametrization from Ref. [10] with 20% [7] uncertainty.

The $\langle m_T \rangle$ of $p$ in isobar collisions in different centralities ranges from 1.24 to 1.36 GeV/$c^2$. Hence, we choose the parametrization $R_{\rm inv} = 0.428\langle dN_{ch}/d\eta\rangle^{1/3} - 0.05$ obtained from $p-p$ correlations functions measured in $\sqrt{s_{\rm NN}} = 5.02$ TeV Pb+Pb collisions in the similar $m_T$ range $1.3-1.4$ GeV/$c^2$. This parameterization is shown in Fig. 2. An uncertainty of 20% is assigned in accordance with Ref. [7]. This parametrization describes the data from $p$+$p$ collisions at $\sqrt{s} = 13$ TeV [9] and Pb+Pb collisions at $\sqrt{s_{\rm NN}} = 2.76$ TeV [11]. For the four centrality intervals considered in this work (0–10%, 10–20%, 20–40%, and 40–80%), the average charged-particle multiplicities for the isobar collisions at $\sqrt{s_{\rm NN}} = 200$ GeV are $\langle dN_{\rm ch}/d\eta\rangle = 276\pm28$, $188\pm18$, $107\pm10$, and $30\pm3$, respectively. Using the ALICE parameterization, the corresponding source radii are $R_{\rm inv} = 2.74\pm0.09$, $2.40\pm0.08$, $1.98\pm0.06$, and $1.28\pm0.04$ fm.

Combining the $R_{\rm inv}$ parameterization and the $B_A$ calculations from Ref. [7], $B_2, B_{3(t)}, B_{3(^3{\rm He})}$, and $B_{3\Lambda}$ can be obtained as a function of $dN_{\rm ch}/d\eta$. Assuming thermal spectral shapes and momentum-independent ratios $B_{3(t)}/B_2^2$ and $B_{3\Lambda}/B_{3(^3{\rm He})}$, the integrated yield ratios are related to the coalescence parameters by:

$$\frac{N_tN_p}{N_d^2} = \frac{3}{4}\times\frac{B_{3(t)}}{B_2^2} \quad (3)$$

$$\frac{N_{^3_\Lambda{\rm H}}/N_{^3{\rm He}}}{N_\Lambda/N_p} = \frac{B_{3\Lambda}}{B_{3(^3{\rm He})}} \quad (4)$$

The wave functions employed in this study for $d$ are identical to those described in Ref. [7]. For the Hulthén wave function [12], the parameters are set to $\alpha = 0.2$ fm$^{-1}$ and $\beta = 1.56$ fm$^{-1}$, corresponding to a root-mean-square radius of 2.13 fm, consistent with experimental measurements [13]. As for the $t$ wave function, we adopt the Gaussian parameterization with a width parameter of $b_t = 1.77$ fm, following Ref. [8], which is consistent with experimental measurements [14]. The $\chi^2$/NDF values obtained from comparisons of the isobar $N_tN_p/N_d^2$ data to calculations using the Gaussian and Hulthén wave functions, as well as other models, are summarized in Tab. III.

TABLE III. Comparison of the isobar $N_tN_p/N_d^2$ data to different model calculations. The quoted $\chi^2$/NDF values are calculated using the four isobar data points, including both experimental and model uncertainties.

| Model | $d$ wave function | $\chi^2$/NDF |
|---|---|---|
| Thermal-FIST | - | 374.3/4 |
| Coalescence-inspired fit | Gaussian | 9.3/4 |
| Hybrid coalescence | Gaussian | 124.6/4 |
| Analytical coalescence | Gaussian | 25.1/4 |
| Analytical coalescence | Hulthén | 14.4/4 |

For $^3$He, we adopt the Gaussian parameterization with a width parameter of $b_{^3{\rm He}} = 1.97$ fm, following Ref. [7], which is consistent with experimental measurements [14]. For the ${}^3_\Lambda$H, our study utilizes both the Gaussian and Congleton wave functions. For the Gaussian wave function, the Gaussian width is chosen such that the root-mean-squared $\Lambda$–deuteron separation $\langle r_{d\Lambda}\rangle$ is $9.8^{+1.1}_{-0.7}$ fm, as inferred from binding-energy measurements [15]. For the Congleton wave function, we use two parameter sets from the literature: one from the original two-body calculation of Congleton [16], and another derived from effective field theory (EFT) calculations incorporating three-body forces [17]. We also perform a $\chi^2$ fit to the isobar $S_3$ data with $Q_\Lambda$ and $\alpha_\Lambda$ treated as free parameters. All parameter sets used in the main manuscript, including the best fit values obtained from the $\chi^2$ fit, together with the corresponding $\langle r_{d\Lambda}\rangle$ values calculated from the wave functions, are summarized in Tab. IV. The $\chi^2$/NDF values obtained from comparisons of the isobar $S_3$ data to calculations using the Gaussian and Congleton wave functions, as well as other models, are summarized in Tab. V.

TABLE IV. Congleton parameter sets employed in the analytical coalescence calculations

| | $Q_\Lambda$ (fm$^{-1}$) | $\alpha_\Lambda$ (fm$^{-1}$) | $\langle r_{d\Lambda}\rangle$ (fm) |
|---|---|---|---|
| 2-body [16] | 1.17 | 0.068 | 11.4 |
| 3-body EFT [17] | 2.5 | 0.068 | 10.9 |
| Best fit | 0.46 | 0.219 | 6.3 |

TABLE V. Comparison of the isobar $S_3$ data to different model calculations. The quoted $\chi^2$/NDF values are calculated using the four isobar data points, including both experimental and model uncertainties. The correlated ${}^3_\Lambda$H branching-ratio uncertainty is treated using a covariance-matrix approach.

| Model | ${}^3_\Lambda$H wave function | $\chi^2$/NDF |
|---|---|---|
| Thermal-FIST | - | 22.9/4 |
| Hybrid coalescence | Gaussian | 13.9/4 |
| Analytical coalescence | Gaussian | 14.9/4 |
| Analytical coalescence | Congleton (2-body) | 1.6/4 |
| Analytical coalescence | Congleton (3-body EFT) | 4.9/4 |
| Analytical coalescence | Congleton (best fit) | 0.9/2 |

## FIT TO ISOBAR $S_3$ DATA USING A GAUSSIAN ${}^3_\Lambda$H WAVE FUNCTION

The Gaussian wave function for ${}^3_\Lambda$H can be written as [7]

$$\Phi_{{}^3_\Lambda\mathrm{H}}(r_{d\Lambda}) \propto e^{-r^2_{d\Lambda}/(3b^2_{d\Lambda})}, \quad (5)$$

where the root-mean-square $d$–$\Lambda$ separation is related to the width parameter by

$$\langle r_{d\Lambda}\rangle = \frac{3}{2} b_{d\Lambda}. \quad (6)$$

Using this wave function, we calculate $S_3$ within the analytical coalescence framework and perform a fit to the isobar data with $b_{d\Lambda}$ as a free parameter. The resulting $\chi^2$ profile as a function of $\langle r_{d\Lambda}\rangle$ is shown in Fig. 3. The fit yields $\langle r_{d\Lambda}\rangle = 4.69^{+0.56}_{-0.44}$ fm, with a minimum $\chi^2$ of 0.9. In contrast, the world-average binding energy, $B_\Lambda = 0.163 \pm 0.036$ MeV, corresponds to $\langle r_{d\Lambda}\rangle = 9.8^{+1.1}_{-0.7}$ fm. When the world-average binding-energy constraint is incorporated into the fit, the minimum $\chi^2$ increases to 18.6, indicated by the $y$-coordinate of the red marker in Fig. 3. The resulting increase of $\Delta\chi^2 = 17.7$ corresponds to a $4.2\,\sigma$ tension between the $S_3$ measurements and the binding-energy constraint within the Wigner-function coalescence framework. This further supports the conclusion that the Gaussian ansatz does not provide an adequate description of the ${}^3_\Lambda$H spatial structure.

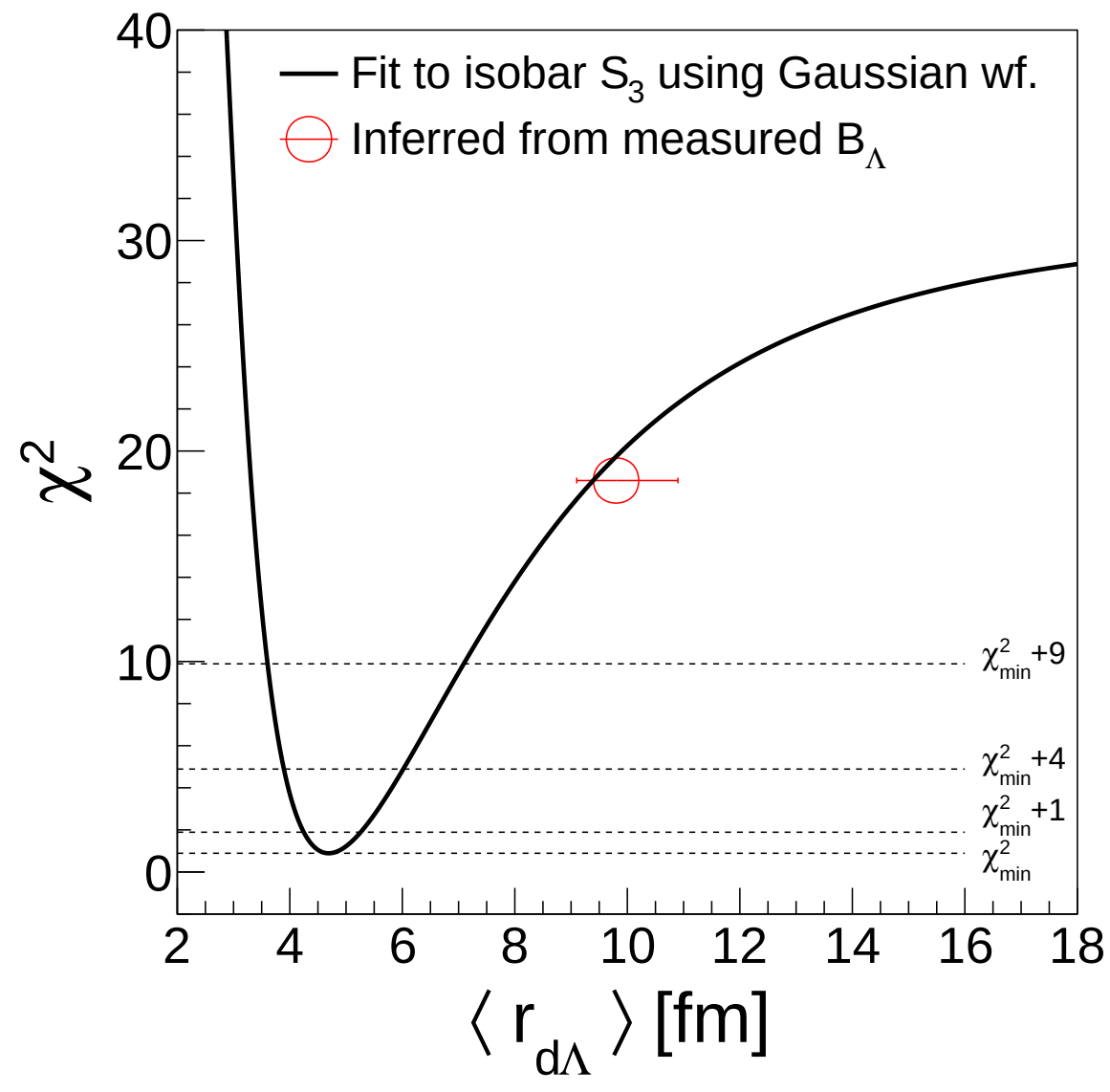


FIG. 3. $\chi^2$ profile as a function of the root-mean-square $d$–$\Lambda$ separation for the Gaussian ${}^3_\Lambda$H wave function. The horizontal lines $\chi^2_{\rm min}+1$, $\chi^2_{\rm min}+4$, and $\chi^2_{\rm min}+9$, corresponding to the $1\sigma$, $2\sigma$, and $3\sigma$ intervals. The $x$-coordinate of the red marker denotes the range implied by the world-average binding energy, while the $y$-coordinate indicates the minimum $\chi^2$ obtained when the world-average binding-energy constraint is included in the fit.



\end{document}